\documentclass[11pt]{article}

\usepackage{bbm}
\usepackage{epsfig}
\usepackage{array}
\usepackage{float}
\usepackage{dsfont}
\usepackage{amstext}
\usepackage{rotating}
\usepackage{a4}
\usepackage{a4wide}
\usepackage{cite}

\def\be{\begin{equation}}
\def\ee{\end{equation}}
\def\gs{\mathrel{
   \rlap{\raise 0.511ex \hbox{$>$}}{\lower 0.511ex \hbox{$\sim$}}}}
\def\ls{\mathrel{
   \rlap{\raise 0.511ex \hbox{$<$}}{\lower 0.511ex \hbox{$\sim$}}}}
\newcommand{\obb}{0\mbox{$\nu\beta\beta$}}
\newcommand{\onbb}{neutrinoless double beta decay }
\newcommand{\ba}{\begin{array}{c}}
\newcommand{\baz}{\begin{array}{cc}}
\newcommand{\barrr}{\begin{array}{rrr}}
\newcommand{\bad}{\begin{array}{ccc}}
\newcommand{\bav}{\begin{array}{cccc}}
\newcommand{\baf}{\begin{array}{ccccc}}
\newcommand{\bea}{\begin{equation} \begin{array}{c}}
\newcommand{\eea}{ \end{array} \end{equation}}
\newcommand{\ea}{\end{array}}
\newcommand{\D}{\displaystyle}
\newcommand{\dms}{\mbox{$\Delta m^2_{\odot}$}}
\newcommand{\dma}{\mbox{$\Delta m^2_{\rm A}$}}
\newcommand{\meff}{\mbox{$\langle m \rangle$}}

\newcommand{\gsim}{\raise0.3ex\hbox{$\;>$\kern-0.75em\raise-1.1ex\hbox{$\sim\;$}}} 
\newcommand{\lsim}{\raise0.3ex\hbox{$\;<$\kern-0.75em\raise-1.1ex\hbox{$\sim\;$}}}

\begin{document}

\title{
\hfill {\small FERMILAB--PUB--08--535--T}\\[-0.1in]
\hfill {\small arXiv: 0812.0436 [hep-ph]} 
\vskip 0.4cm
\bf 
Comparing Trimaximal Mixing and Its Variants 
with Deviations from Tri-bimaximal Mixing}
\author{
Carl H.~Albright$^{a,b}$\thanks{email: 
\tt albright@fnal.gov}~\mbox{ 
},~~Werner Rodejohann$^c$\thanks{email: 
\tt werner.rodejohann@mpi-hd.mpg.de}
\\\\
{\normalsize \it$^a$Department of Physics, Northern Illinois University,}\\
{\normalsize \it DeKalb, Illinois 60115, USA}\\ \\ 
{\normalsize \it$^b$Fermi National Accelerator Laboratory,}\\
{\normalsize \it Batavia, Illinois 60510, USA}\\ \\ 
{\normalsize \it$^c$Max--Planck--Institut f\"ur Kernphysik,}\\
{\normalsize \it  Postfach 103980, D--69029 Heidelberg, Germany} 
}
\date{}
\maketitle
\thispagestyle{empty}
\vspace{-0.8cm}
\begin{abstract}
\noindent  
We analyze in detail the predictions of ``trimaximal'' neutrino
mixing, which is defined by a mixing matrix with identical second
column elements. This column is therefore identical to the second
column in the case of tri-bimaximal mixing. We also generalize 
trimaximal mixing by assuming that the other rows and columns of 
the mixing matrix individually can have the same forms as for 
tri-bimaximal mixing. The phenomenology of these 
alternative scenarios and their mixing angle and CP phase correlations
are studied.  
We emphasize how trimaximal mixing scenarios can be
distinguished experimentally from broken tri-bimaximal mixing.  

\end{abstract}

\newpage

\section{Introduction}

\indent Ten years after observation of the depletion of 
atmospheric muon-neutrinos was established by the 
SuperKamiokande collaboration 
\cite{S-K1998}, our knowledge of the neutrino 
oscillation parameters has been 
noticeably sharpened by the ensuing atmospheric \cite{atm}, 
solar \cite{sol}, reactor \cite{reactor}, and 
long-baseline \cite{lbl} neutrino experiments.  
Recent global analyses \cite{concha,thomas,bari} 
limit the oscillation parameters to the $1\sigma$ and 
$3\sigma$ ranges determined by Fogli {\it et al.}, 
for example, to be \cite{bari}
\begin{equation}
\begin{array}{rcl}
\Delta m^2_{32} &=& 
\left(2.39^{+0.11,\,0.42}_{-0.08,\,0.33}\right)
	\times 10^{-3}\ {\rm eV^2}\,,\\[0.24cm]
\Delta m^2_{21} &=& \left(7.67^{+0.16,\,0.52}_{-0.19,\,0.53}\right)
	\times 10^{-5}\ {\rm eV^2}\,,\\[0.24cm]
\sin^2 \theta_{23} &=& 0.466^{+0.073,\,0.178}_{-0.058,\,0.135} \,,\\[0.24cm]
\sin^2 \theta_{12} &=& 0.312^{+0.019,\,0.063}_{-0.018,\,0.049} \,,\\
\end{array}
\label{eq:data}
\end{equation}	

\noindent while the reactor neutrino oscillation angle 
remains more uncertain. 
In a recent analysis \cite{bari} it is weakly ($1.6\sigma$) 
constrained to be non-zero according to 
(see also Ref.~\cite{other0})
\begin{equation}
  \sin^2 \theta_{13} = 0.016 \pm 0.010\ (\le 0.046) \, .\\
\label{eq:data13}
\end{equation}

These mixing angles help to specify the 
Pontecorvo-Maki-Nakagawa-Sakata (PMNS)
mixing matrix defined in the standard convention \cite{PDG} by 
\be \label{eq:U}
U_{\rm PMNS} = \left( \bad 
c_{12}  \, c_{13} 
& s_{12} \, c_{13} 
& s_{13} \, e^{-i \delta}  \\ 
-s_{12} \, c_{23} 
- c_{12} \, s_{23} \, 
s_{13}  \, e^{i \delta} 
& c_{12} \, c_{23} - 
s_{12} \, s_{23} \, s_{13} 
\, e^{i \delta} 
& s_{23}  \, c_{13}  \\ 
s_{12}   \, s_{23} - c_{12} 
\, c_{23}  \, s_{13} \, e^{i \delta} & 
- c_{12} \, s_{23} 
- s_{12} \, c_{23} \, 
s_{13} \, e^{i \delta} 
& c_{23}  \, c_{13}  
\ea   
\right) \,,
\ee
where $c_{ij} = \cos \theta_{ij},\ s_{ij} = 
\sin \theta_{ij}$ with $\delta$ the unknown CP-violating 
Dirac phase.  
Harrison, Perkins, and Scott first emphasized that the 
experimentally obtained mixing matrix is close to the simple 
tri-bimaximal mixing (TBM) form where \cite{tri} 
\be
U_{\rm TBM} = 
\left( 
\bad 
\sqrt{\frac 23} & \sqrt{\frac 13} & 0 \\
-\sqrt{\frac 16} &  \sqrt{\frac 13} &  -\sqrt{\frac 12} \\ 
-\sqrt{\frac 16} &  \sqrt{\frac 13} &  \sqrt{\frac 12}
\label{eq:UTBM}
\ea
\right)\,.
\ee

\noindent The column vectors are just the eigenvectors of 
the three diagonal $U(3)$ operators.  For exact tri-bimaximal 
mixing, the mixing angles are\footnote{The experimental results 
are so close to TBM that parametrizations of the PMNS 
matrix with TBM as the starting point have been proposed 
\cite{triM}.} 
\begin{equation}
  \sin^2 \theta_{12} = \frac{1}{3}\,,\quad 
\sin^2 \theta_{23} = \frac{1}{2}\,,
  \quad \sin^2 \theta_{13} = 0\,,
\label{eq:TBM}
\end{equation}

\noindent which are seen to be close to the present 
data quoted above in Eq.~(\ref{eq:data}).

In a top-down approach, the charged lepton mass 
matrix, $m_\ell$, and 
light Majorana neutrino mass matrix, $m_\nu$, are 
specified in some model 
with particular family and flavor symmetries, often 
by invoking the seesaw mechanism.  The two mass matrices 
are diagonalized by two unitary transformations such that 
\be
\label{eq:topdown}
\begin{array}{rcl}
   U^\dagger_\ell \, m_\ell^\dagger m_\ell \, U_\ell &=& 
            {\rm diag}(m^2_e,\, m^2_\mu,\, m^2_\tau) \,,\\[0.2cm]
   U^T_\nu \, m_\nu \, U_\nu &=& 
m^{\rm diag}_\nu = {\rm diag}(m_1,\, m_2,\, m_3) \, .\\
\end{array}
\ee

\noindent  The PMNS mixing matrix then follows from 
$U_{\rm PMNS} \, P = U^{\dagger}_\ell \, U_\nu$, where the 
diagonal Majorana phase matrix is given by 
$P = {\rm diag}(1, \, e^{i\alpha}, \, e^{i\beta})$. 
Note that without this matrix the phase transformation required on 
the right side of $U_{\rm PMNS}$ 
(and hence on $U_\nu$ itself) to bring it 
into the conventional phase structure of Eq.~(\ref{eq:U}) 
is not possible, when one demands real positive 
diagonal neutrino mass entries in Eq.~(\ref{eq:topdown}). 
The presence of $P$ allows one to compensate for the 
phase transformation on the right side of $U_{\rm PMNS}$ 
without altering $U_\nu$.
In this top-down approach one can then compare the 
$U_{\rm PMNS}$ obtained with $U_{\rm TBM}$.

Instead, one might employ a bottom-up procedure 
to identify the $\mu$--$\tau$ symmetric neutrino mass matrix 
in the flavor basis as the most general one giving rise 
to tri-bimaximal mixing,
\be \label{eq:mnutbm}
(m_\nu)_{\rm TBM} = 
U_{\rm TBM}^\ast\,P^\ast\, m_\nu^{\rm diag}\,P^\dagger\, 
U_{\rm TBM}^\dagger = 
\left(
\bad 
A & B & B \\[0.2cm]
\cdot & \frac{1}{2} (A + B + D) & \frac{1}{2} (A + B - D)\\[0.2cm]
\cdot & \cdot & \frac{1}{2} (A + B + D)
\ea 
\right)\,.
\ee
The parameters $A,B,D$ are in general complex and functions 
of the neutrino masses and Majorana phases:
\be \D 
A = \frac 13 \left(2 \, m_1 + m_2 \, e^{-2i\alpha} \right) \,,~~
B = \frac 13 \left(m_2 \, e^{-2i\alpha} - m_1 \right) \,,~~
\D D = m_3 \, e^{-2i\beta} \,.
\ee
Another way to write the mass matrix is to decompose it 
in terms of the three individual masses. 
\be \label{eq:mnuind}
(m_\nu)_{\rm TBM} = 
\frac{m_1}{6} \, 
\left( 
\bad 
4 & -2 & -2 \\
\cdot & 1 & 1 \\
\cdot & \cdot & 1 
\ea 
\right) 
+ \frac{m_2 \, e^{-2i\alpha}}{3} \, 
\left( 
\bad 
1 & 1 & 1 \\
\cdot & 1 & 1 \\
\cdot & \cdot & 1 
\ea 
\right) + 
\frac{m_3 \, e^{-2i\beta}}{2} \, 
\left( 
\bad 
0 & 0 & 0 \\
\cdot & 1 & -1 \\
\cdot & \cdot & 1 
\ea 
\right) .
\ee 

Since observable departures from exact tri-bimaximal 
mixing are expected, it is of interest to study how 
deviations may arise.  In a recent paper the 
authors linearly perturbed the matrix elements 
in Eq.~(\ref{eq:mnutbm}) in order to observe how large 
the departures of the mixing angles from their 
tri-bimaximal values in Eq.~(\ref{eq:TBM}) can be \cite{AR}.  
Allowing up to 20\% deviations in the matrix elements 
in the normal hierarchy case results in non-zero values 
for $\sin^2 \theta_{13}$ up to 0.001, for example. 
We argued that larger deviations in connection 
with a normal mass hierarchy would signal that 
the apparent nearly tri-bimaximal mixing should be 
considered accidental in nature, rather than the 
result of a softly-broken symmetry.

In this paper we study other mixing scenarios which
deviate from tri-bimaximal mixing by leaving only one of 
the columns or one of the rows invariant. We shall 
refer to such scenarios as ``generalized trimaximal 
mixing.''  The term ``trimaximal mixing'' \cite{TM3} was 
originally introduced to describe a mixing matrix in 
which only the second column of Eq.~(\ref{eq:UTBM}) 
remains invariant with the absolute value of 
every element of that column equal to $1/\sqrt{3}$. 
To distinguish the different versions, we label the 
original version of trimaximal mixing considered in 
\cite{TM0,TM1,TM2,TM3,TM4} as TM$_2$, i.e., 
\be \label{eq:TM}
\mbox{TM}_2 :~~
\left|U_{\alpha 2} \right|^2 = \frac 13\,,~~~~ \forall 
\, \alpha = e, \,\mu,\, \tau\,.
\ee
We shall also study the effects of allowing the other two 
columns to remain independently invariant under 
tri-bimaximal mixing perturbations and label 
them ${\rm TM}_1$ and ${\rm TM}_3$, respectively. It is also 
of interest to allow 
one of the rows to remain invariant for which we adopt 
the labeling ${\rm TM}^i \, ,\ {\rm with}\ 
i = 1,\ 2,\ {\rm or}\ 3$. In total there are six possibilities.  
By examining these variations we shall learn that the 
three mixing angles and the 
CP-violating Dirac phase are not all independent and 
that restricted ranges of the deviations are imposed by 
the present mixing data. In fact, keeping a row or a column of 
$U_{\rm PMNS}$ fixed leads to a 2-parameter scenario, i.e., 
two of the four mixing observables in 
$U_{\rm PMNS}$ are determined. In four of the six trimaximal 
variants we find an interesting, characteristic 
and testable correlation between the mixing parameters. 
Future more precise data will allow one to test some of the 
trimaximal variations considered here. In particular, we 
stress that while TM$_2$ allows naturally for
non-zero $\theta_{13}$, non-maximal $\theta_{23}$ 
and for $\sin^2 \theta_{12} \neq \frac 13$, it predicts that 
$\sin^2 \theta_{12} \ge \frac 13$, while the current best-fit points
all lie at $\sin^2 \theta_{12} \le \frac 13$. 
Some of the alternative trimaximal scenarios can accommodate 
this, while still allowing for non-zero $\theta_{13}$. 

We would like to stress that in the present paper 
not merely deviations from tri-bimaximal mixing are discussed, 
but also novel and testable mixing scenarios are presented. 
Though tri-bimaximal mixing dominates the current 
phenomenological and theoretical discussion in neutrino physics, 
alternative proposals are surely of interest, and here we study one
possible avenue.\\ 

In Sect.~\ref{sec:AR} we summarize the results obtained 
previously for broken tri-bimaximal mixing in \cite{AR}. 
In Sect.~\ref{sec:TM} we derive the results 
for the originally proposed trimaximal mixing and for the 
variant forms of trimaximal mixing in 
Sect.~\ref{sec:TM1}.  A comparison of the results and conclusions 
are presented in Sect.~\ref{sec:concl}.

\section{\label{sec:AR}Deviations from Tri-bimaximal Mixing}

In \cite{AR} we raised the issue of deviations in the 
mixing observables from tri-bimaximal mixing when the mass 
matrix is perturbed.  Our strategy was to perturb every 
entry of the mass matrix with a small complex
parameter $\epsilon_i$, i.e:
\be \label{eq:mnu_dev}
m_\nu =   
\left(
\bad 
A \, (1 + \epsilon_1) & B \, (1 + \epsilon_2) 
& B \, (1 + \epsilon_3)\\[0.2cm]
\cdot & \frac{1}{2} (A + B + D) \, (1 + \epsilon_4) 
& \frac{1}{2} (A + B - D) \, (1 + \epsilon_5)\\[0.2cm]
\cdot & \cdot & \frac{1}{2} (A + B + D) 
\, (1 + \epsilon_6)
\ea 
\right)\,,
\ee
where the complex perturbation parameters were taken to 
be $|\epsilon_i| \le 0.2$ for $i = 1 - 6$ with their 
phases $\phi_i$ allowed to lie 
between zero and $2\pi$. The results obtained for the 
oscillation parameters depend on the neutrino mass values 
and ordering. In case of a normal
hierarchy ($m_3 \gg m_2 > m_1$), the maximal expected values are 
\be
\mbox{NH:\qquad} |U_{e3}| \ls \frac 23 \, \sqrt{\frac\dms\dma} 
\, |\epsilon| \simeq 0.027 \,,~\mbox{ and } 
\left|\frac 12 - \sin^2 \theta_{23} \right| 
\ls \frac 12 \, |\epsilon| \simeq 0.1\,,
\ee
where $|\epsilon|$ denotes the absolute value of one of the six
independent breaking parameters in Eq.~(\ref{eq:mnu_dev}). 
Fig.~\ref{fig:m1} shows for a maximal 20\% deviation the 
parameter $|U_{e3}|^2$ as a function of the smallest 
neutrino mass $m_1$. Note that the maximal 
value of $|U_{e3}|$ depends linearly on $\epsilon$. Hence, if 
the deviation is allowed to range up to 50\% 
for the normal hierarchy case, one finds that 
$|U_{e3}|^2$ can reach a maximum value of roughly 
0.005 for $m_1 \ls 4$ meV. 
Even for a 
normal hierarchy it turns out that $\sin^2 \theta_{12}$ can take
values anywhere in its allowed range. 
The latter is also true in case of an inverted hierarchy 
($m_2 \simeq m_1 \gg m_3$), for which one furthermore finds 
that the other mixing observables are bounded from above by 
\be
\mbox{IH:\qquad} |U_{e3}| \ls 
\frac 13 |\epsilon| \sqrt{ \frac{8}{9} 
+ \frac{16}{3} \, \frac{m_3}{\sqrt{\dma}} } 
\simeq  0.12 \,,~\mbox{ and } 
\left|\frac 12 - \sin^2 \theta_{23} \right| 
\ls \frac 89 \, |\epsilon| \simeq 0.18\,.
\ee
Almost all of the $3\sigma$ range can be covered. 
The interesting accessible range for $|U_{e3}|^2$ as 
a function of the smallest neutrino mass $m_3$ is 
also plotted in Fig.~\ref{fig:m1}. 
If neutrinos are quasi-degenerate, then the fully allowed 
parameter space can be covered.

In Ref.~\cite{AR} we also presented the results for some predictive 
$SO(10)$ symmetric Grand Unified models exhibiting a 
normal mass hierarchy. 
There we found that $|U_{e3}|^2 \gs 2 \times 10^{-3}$, while 
$|\frac 12 - \sin^2 \theta_{23}| \gs 0.07$ with $\sin^2 \theta_{12}$ 
typically $< \frac{1}{3}$.  Hence one can distinguish the 
results of these models from softly-broken tri-bimaximal mixing once 
the value of $|U_{e3}|^2$ and the neutrino mass hierarchy is known.

\section{\label{sec:TM}Trimaximal Mixing Deviations from TBM}

We now turn to study the deviations from tri-bimaximal 
mixing which can 
arise with the less restrictive symmetry referred to as 
trimaximal mixing, defined by Eq.~(\ref{eq:TM}):
\be 
\mbox{TM}_2:~~
\left(
\ba 
|U_{e2}|^2 \\
|U_{\mu 2}|^2 \\
|U_{\tau 2}|^2
\ea
\right) = 
\left(\ba 1/3 \\ 1/3 \\ 1/3 \\ \ea \right) \,.
\ee
This is the original version of trimaximality, for which 
also a model based on the $\Delta(27)$ flavor symmetry was proposed
\cite{TM3,TM4}. 
From the condition $|U_{e 2}|^2 = \frac 13 $ it follows that 
\be \label{eq:TM1}
\sin^2 \theta_{12} = \frac 13 \, \frac{1}{1 - |U_{e3}|^2} \ge
\frac 13 \,.
\ee
Note that ${\rm TM}_2$ predicts 
$\sin^2 \theta_{12} \ge \frac 13$, to be
compared with the current best-fit values and 
$1\sigma$ ranges of Ref.~\cite{bari}, 
$\sin^2 \theta_{12} = 0.312_{-0.018}^{+0.019}$,  
and Ref.~\cite{thomas}: 
$\sin^2 \theta_{12} = 0.304_{-0.016}^{+0.022}$. 
Inserting the range $|U_{e3}|^2 = 0.016 \pm 0.010$ would give 
$\sin^2 \theta_{12} = 0.339 \pm 0.003$.  

The second prediction of trimaximal mixing can be obtained from 
$|U_{\mu2}|^2 = \frac 13$ and is  
\bea
\label{eq:TM2} \D
\cos \delta \, \tan 2 \theta_{23} = \frac{2 \, \cos \theta_{13} \, 
\cot 2 \theta_{13}}{\sqrt{2 - 3 \, \sin^2 \theta_{13}}} 
= \frac{1 - 2 \, |U_{e3}|^2}{|U_{e3}| \, 
\sqrt{2 - 3 \, |U_{e3}|^2}}
    \\[0.3cm] \D \simeq   
\frac{1}{\sqrt{2}} \frac{1}{|U_{e3}|}
\left(1 - \frac{5}{4}\, |U_{e3}|^2 
    + {\cal O}(|U_{e3}|^4)\right)\,.
\eea
We observe that the three mixing angles are not independent 
but related as above.  Because $|U_{\tau2}|^2$ is related 
by unitarity with $|U_{e2}|^2$ 
and $|U_{\mu2}|^2$, there is no third independent condition. 

For $\delta \neq \pi/2$ (or $\delta \neq 3\pi/2$) and 
$\theta_{13}$ non-zero, $\theta_{23}$ 
is non-maximal.  On the other hand, for 
maximal CP violation ($\delta = \pi/2$ or $3\pi/2$) it follows that 
$\sin^2 \theta_{23} = \frac 12$, independent of $|U_{e3}|$. 
We can use the expressions for $\sin^ 2 \theta_{12}$ and $\tan 2
\theta_{23}$ to evaluate the Jarlskog invariant for leptonic CP
violation, which in general reads $J_{\rm CP} = 
{\rm Im}(U_{e1}^\ast \, U_{\mu 3}^\ast \, U_{e3} \, U_{\mu 1}) = 
\frac 18 
\, \sin 2 \theta_{13} \, \cos \theta_{13} \, \sin 2 \theta_{23} \, 
\sin 2 \theta_{12} \sin \delta$. We find 
\be \label{eq:Jcp}
J_{\rm CP} \simeq \frac{\sin 2 \delta}{6 \sqrt{2} \, \sqrt{\cos^2
\delta}} \, |U_{e3}| 
\,,
\ee
where we have expanded the lengthy exact equation, which is an odd 
function of $\theta_{13}$.

\begin{figure}[ht]
\begin{center}
\epsfig{file=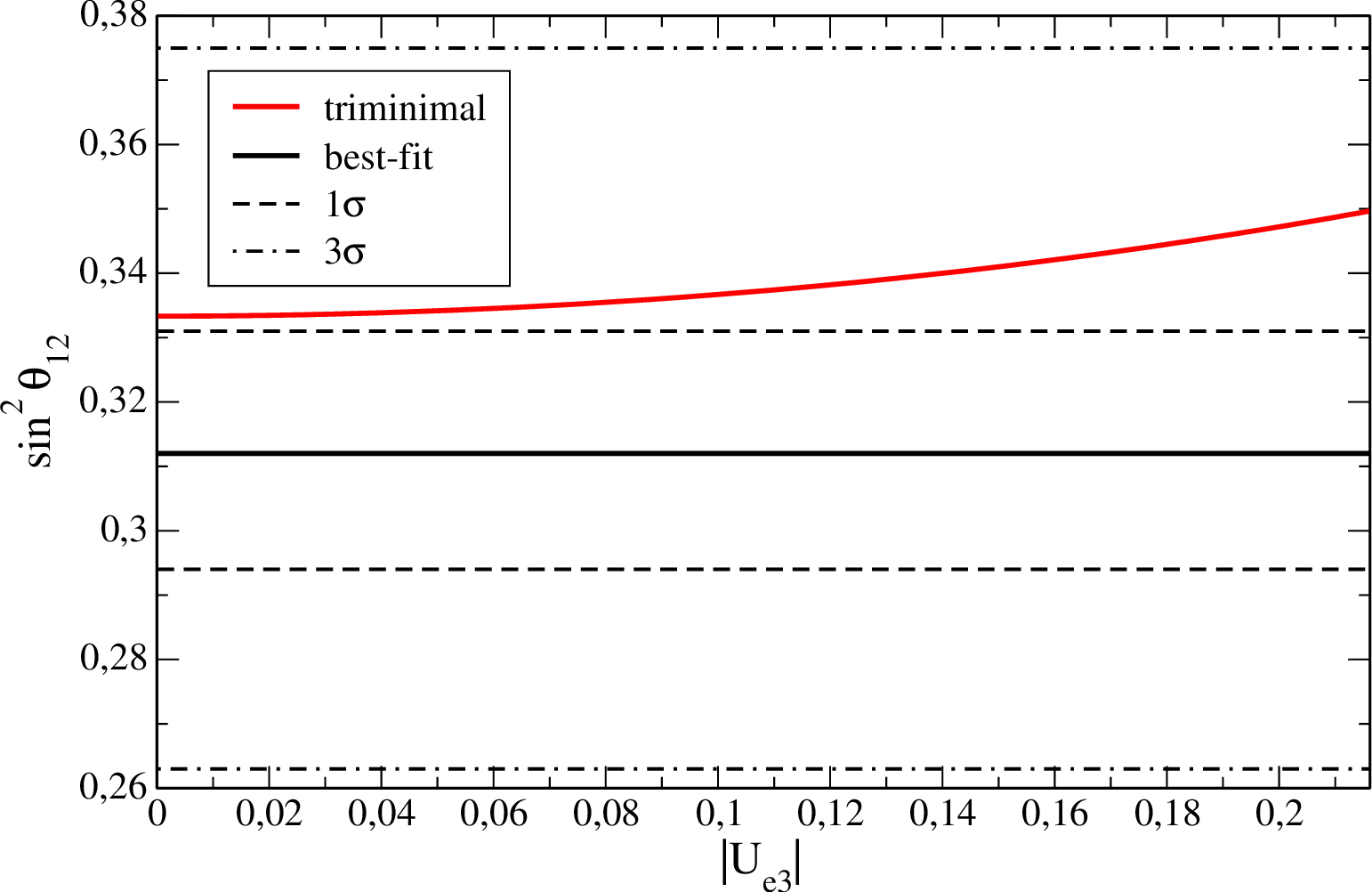,width=10cm,height=7cm}
\epsfig{file=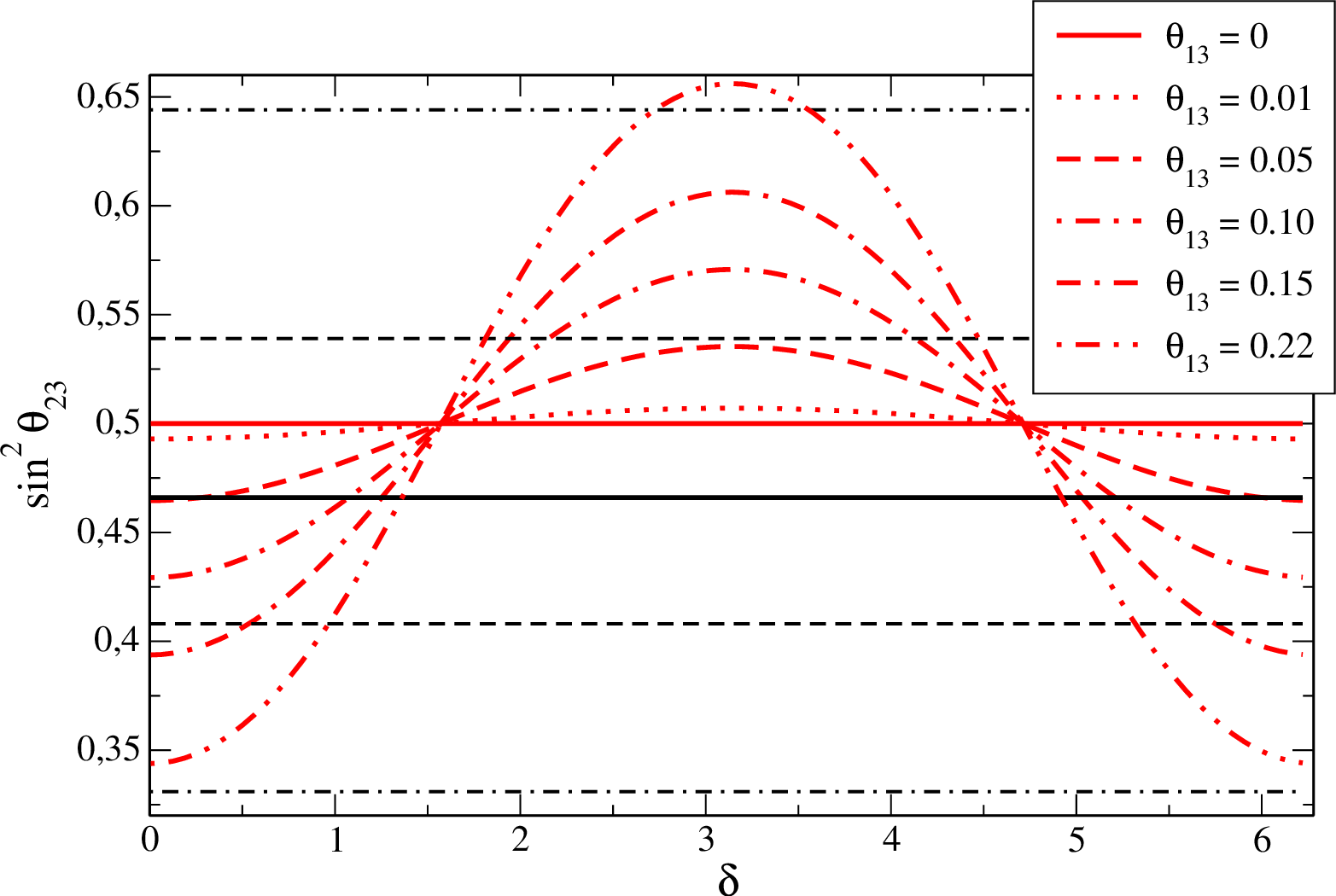,width=10cm,height=7cm}
\epsfig{file=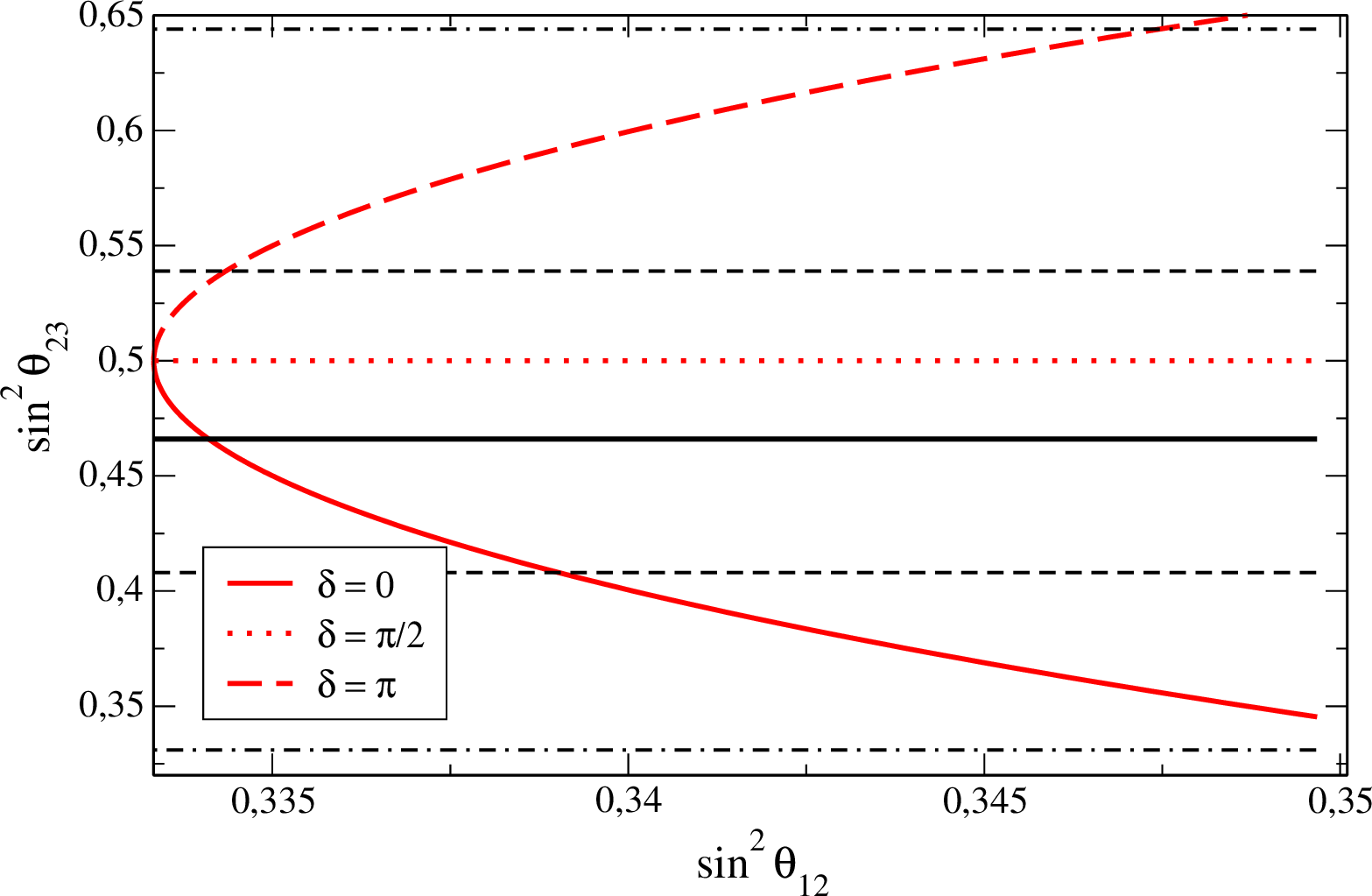,width=10cm,height=7cm}
\end{center}
\vspace*{-0.2in}
\caption{\label{fig:TMt12}Phenomenology of exact TM$_2$ mixing. 
Shown are the solar neutrino parameter $\sin^2
\theta_{12}$ against $|U_{e3}|$, the atmospheric neutrino 
parameter $\sin^2 \theta_{23}$ against $\delta$ for 
different values of $|U_{e3}|$ and $\sin^2
\theta_{12}$ against $\sin^2 \theta_{23}$. 
Also given are the current best-fit value and the $1\sigma$ as well as
$3\sigma$ ranges from a global fit \cite{bari}.  }
\end{figure}

In Fig.~\ref{fig:TMt12} we show plots of $\sin^2 \theta_{12}$ and 
$\sin^ 2 \theta_{23}$ as functions of $|U_{e3}|$ and 
$\delta$, respectively. As can be seen, solar neutrino mixing 
is well within the allowed $2\sigma$ range, and the possible 
deviation from maximal atmospheric 
neutrino mixing is largest for CP conserving values of $\delta$,  
grows with $|U_{e3}|$, and can exceed the $3\sigma$ range. 
Also displayed in Fig.~\ref{fig:TMt12} is a plot of 
$\sin^2 \theta_{12}$ vs.~$\sin^ 2 \theta_{23}$, 
when $|U_{e3}|$ is allowed to vary. It is evident that 
the deviation from maximal atmospheric mixing can be larger than 
the deviation from $\sin^2 \theta_{12} = \frac 13$. 
Indeed, from 
$\theta_{23} = \pi/4 - \epsilon$, one obtains the leading order 
expressions $\tan 2 \theta_{23} \simeq \frac{1}{2\epsilon}$ and  
$\sin^2 \theta_{23} \simeq \frac 12 - \epsilon$. By taking 
only the first order term on the RHS of Eq.~(\ref{eq:TM2}) 
one obtains the relation 
\be
\left(\frac 12 - \sin^2 \theta_{23}\right)^2 
\simeq \frac{\cos^2 \delta }{2} \, |U_{e3}|^2 
\simeq \frac{3 \, \cos^2 \delta }{2} 
\left( \sin^ 2 \theta_{12} - \frac 13 \right)\,.
\ee
This expression shows that the deviation from maximal
atmospheric neutrino mixing can be stronger than the deviation from 
$\sin^2 \theta_{12} = \frac 13$. In general, the 
deviation in $|U_{e3}|^2$ obtained here can be considerably 
larger than that entertained in Sect.~\ref{sec:AR}.

The trimaximal mixing matrix can be parametrized by the application
of a general 13-rotation from the right \cite{TM2}
\be
\label{eq:TM2approx}
U_{\rm TM_2} = U_{\rm TBM} \, R_{13}(\theta; \psi)  \,,
~~\mbox{where } 
R_{13}(\theta; \psi) = 
\left( 
\bad 
\cos \theta & 0 & \sin \theta \, e^{-i \psi} \\
0 & 1 & 0 \\
-\sin \theta \, e^{i \psi} & 0 & \cos \theta 
\ea
\right)\,.
\ee
It is easy to see that $|U_{e3}|^2 = \frac 23 \, \sin^2 \theta$ and 
$\sin^2 \theta_{12} = 1/(3 - 2 \, \sin^2 \theta)$, which is equal to 
$\frac 13 /(1 - |U_{e3}|^2)$ as before. We find furthermore that
\be
\sin^2 \theta_{23} = \frac 12 + \frac{1}{2\sqrt{3}} \, 
\frac{\sin 2 \theta \, \cos \psi}{1 - |U_{e3}|^2}
\mbox{ and } J_{\rm CP} = \frac{\sin 2 \theta \, \sin \psi}
{6\sqrt{3}}\,,
\ee
where 
$\sin 2 \theta = \sqrt{6} \, |U_{e3}| \, \sqrt{1 - \frac32 \,
|U_{e3}|^2}$. 
The CP phase $\delta$ in this parameterization is 
related to the phase $\psi$. 

By using the above form for $U_{\rm TM_2}$ in Eq. 
(\ref{eq:TM2approx}), we can obtain the corresponding neutrino 
mass matrix in the lepton flavor basis from 
\[ 
\label{eq:TM2matrix}
(m_\nu)_{\rm TM_2} = U^\ast_{\rm TM_2}\,P^\ast\, m^{\rm diag}_\nu \, 
      P^\dagger\, U^\dagger_{\rm TM_2} 
= \left( \begin{array}{ccc} A & B + C & B - C\\ 
      \cdot & \frac{1}{2}(A + B + D - 2 \, C) 
& \frac{1}{2}(A + B - D)\\
      \cdot & \cdot & \frac{1}{2}(A + B + D + 2 \, C)\\
      \end{array}\right)\,
\] 
where we identify (with $c_\theta = \cos \theta$ 
and $s_\theta = \sin \theta$)
\[ 
\label{eq:TM2elem}
\begin{array}{rlrl} 
  A &= \frac{1}{3} 
\left(2\,m_1\,c^2_\theta + 2\,m_3 \, s^2_\theta \, 
      e^{2i(\psi-\beta)} 
+ m_2 \, e^{-2i\alpha} \right) , \quad& 
  C &= \frac{1}{\sqrt{3}} 
  \left(m_1 e^{-i\psi} 
- m_3 \, e^{i(\psi-2\beta)}\right) s_\theta \, c_\theta \,,
    \\
  B &= \frac{1}{3}  \left(m_2 \, e^{-2i\alpha} - m_1 \, c^2_\theta - 
      m_3 \, s^2_\theta \, e^{2i(\psi-\beta)}  \right), \qquad&
  D &= m_3 e^{-2i\beta} c^2_\theta 
+ m_1 e^{-2i\psi} s^2_\theta\,.\\
\end{array}
\]
It is clear from this mass matrix that the additional 
$C$ terms break the original $\mu$--$\tau$ symmetry 
present with tri-bimaximal mixing in a well-defined way. 
Note that $C$ vanishes for $\theta = 0$ and that in this case 
$(m_\nu)_{\rm TBM}$ from Eq.~(\ref{eq:mnutbm}) is recovered. 
Interestingly, if we decompose the mass matrix for $\rm TM_2$ 
in terms of the individual neutrino masses, as done for TBM 
in Eq.~(\ref{eq:mnuind}), we find that $m_2$ is multiplied with 
the same flavor-democratic matrix as in Eq.~(\ref{eq:mnuind}). 
The other two masses are multiplied now with 
more complicated matrices, having entries depending 
on the angle $\theta$. 

Since we can now trade $\theta_{12}$ for 
$\theta_{13}$ with the use of Eq.~(\ref{eq:TM1}), it is 
possible to obtain a simple value for the 
effective mass $\langle m_{ee} \rangle$ governing 
\onbb (\obb). With inverted hierarchy ($m_2 \simeq m_1 \gg m_3$) 
the general result is $\meff = c_{13}^2 \, \sqrt{\dma} \, 
\sqrt{1 - \sin^ 2 2 \theta_{12} \,
\sin^ 2 \alpha}$. 
In case of TM$_2$ mixing we then find
\be
\meff \simeq \sqrt{\dma} \, \left[ 
\sqrt{1 - \frac 89 \, \sin^2 \alpha} - 
\frac{1 - \frac 23 \, \sin^2 \alpha}{\sqrt{1 - \frac 89 \, \sin^2
\alpha}} \, |U_{e3}|^2
\right]\,.
\ee
The $\frac 89$ in the first term is of course the value of 
$\sin^ 2 2 \theta_{12}$ in case of tri-bimaximal mixing. 

\section{\label{sec:TM1}Variant Trimaximal Mixing Scenarios}

\begin{figure}[ht]
\begin{center}
\epsfig{file=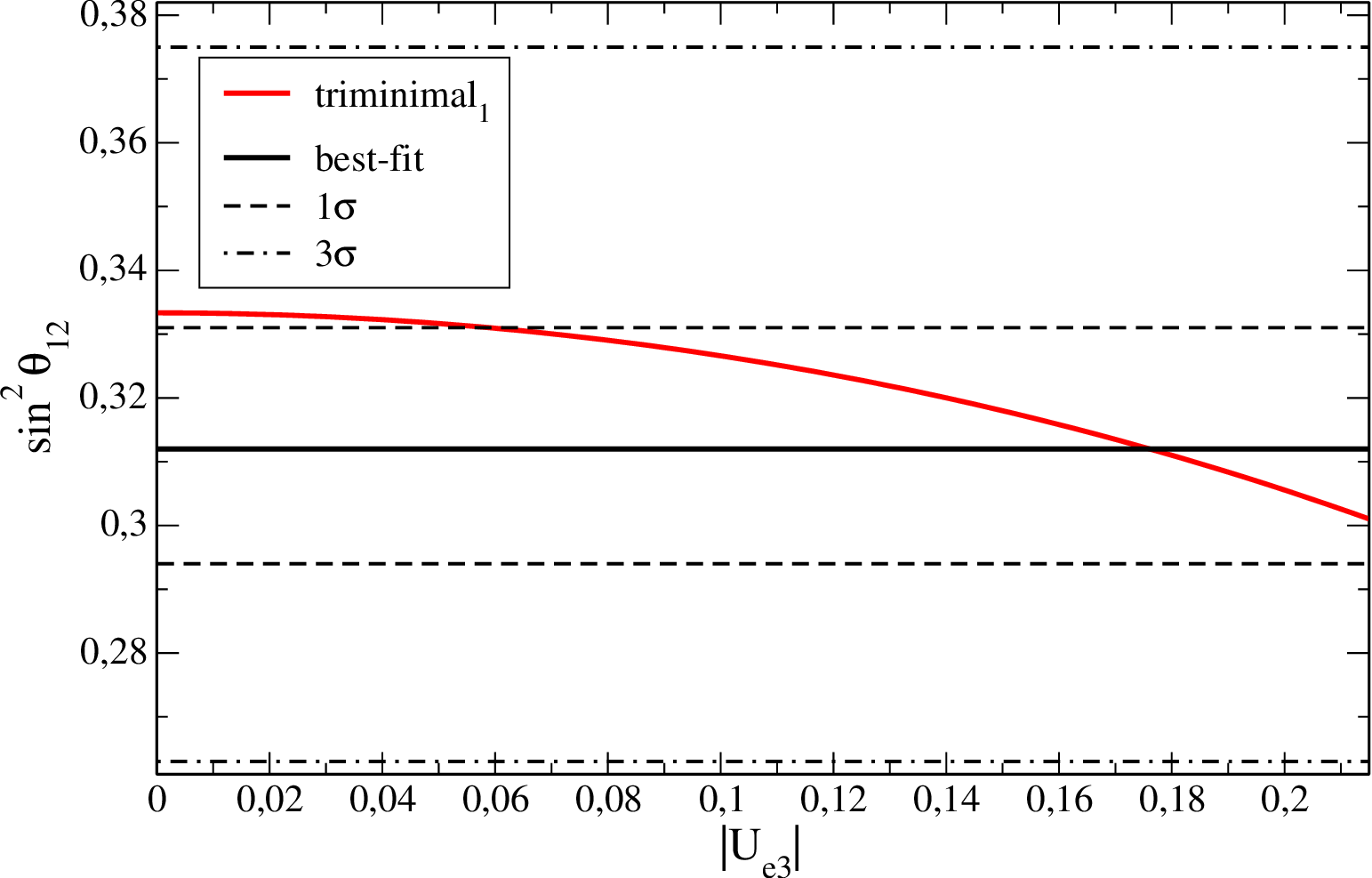,width=10cm,height=7cm}
\epsfig{file=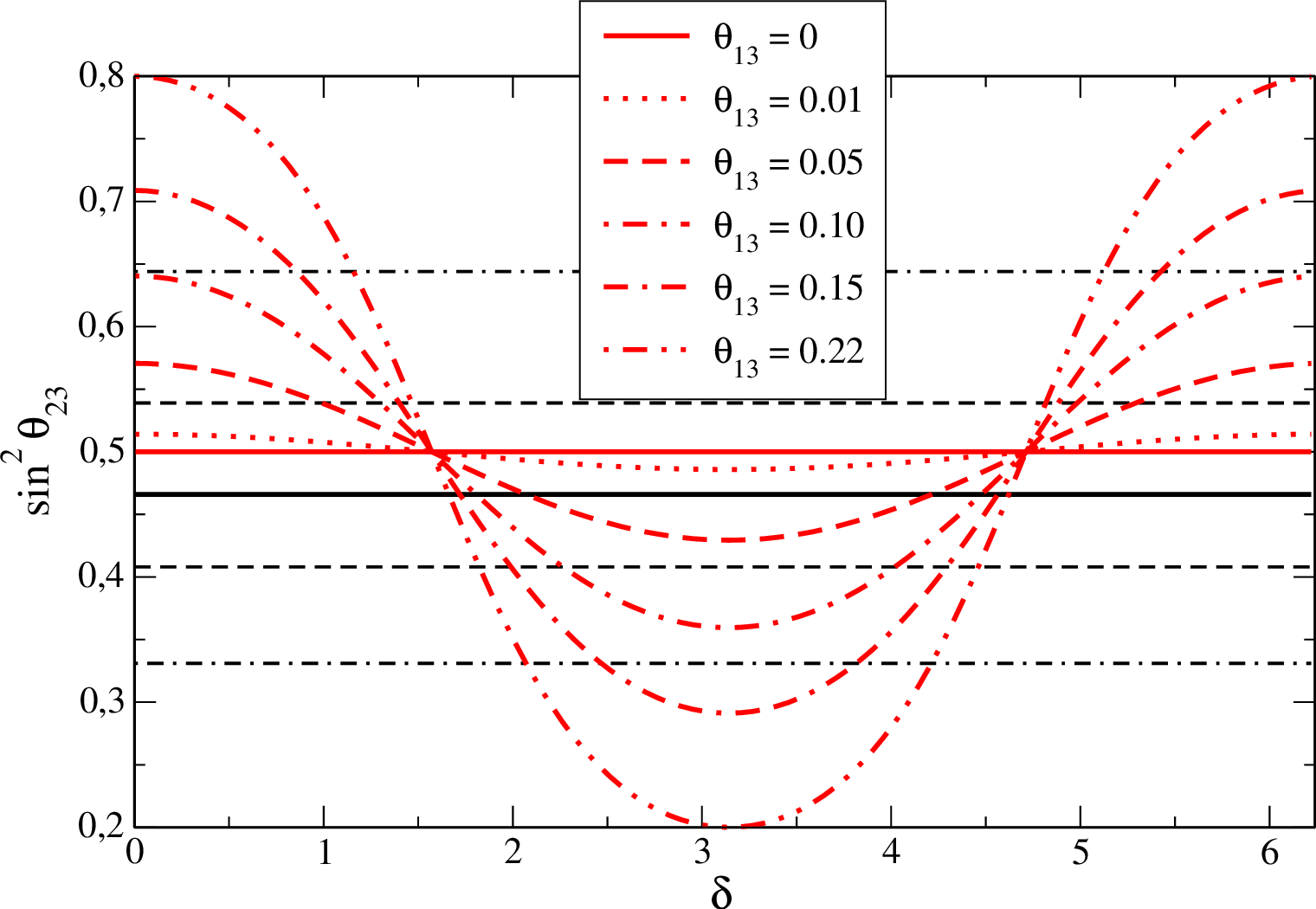,width=10cm,height=7cm}
\epsfig{file=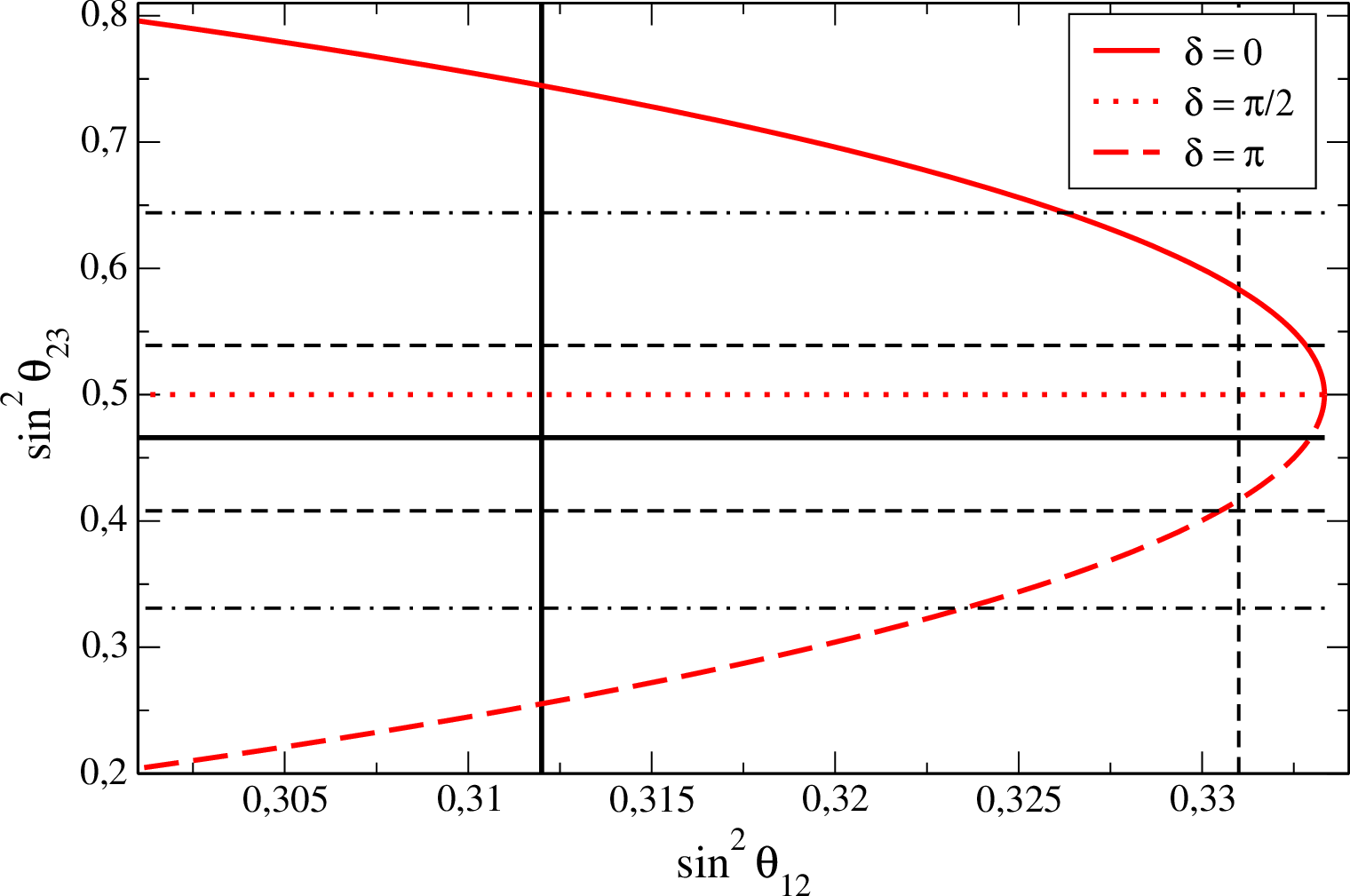,width=10cm,height=7cm}
\caption{\label{fig:TMt12a}
Phenomenology of exact TM$_1$ mixing. 
Shown are the solar neutrino parameter $\sin^2
\theta_{12}$ against $|U_{e3}|$, the atmospheric neutrino 
parameter $\sin^2 \theta_{23}$ against $\delta$ for 
different values of $|U_{e3}|$ and $\sin^2
\theta_{12}$ against $\sin^2 \theta_{23}$. 
Also given are the current best-fit value and the $1\sigma$ as well as
$3\sigma$ ranges from a global fit \cite{bari}.} 
\end{center}
\end{figure}

We have seen that the prediction of trimaximality implies 
$\sin^2 \theta_{12} \ge \frac 13$, while the best-fit points are all
below $\frac 13$. If this trend continues, then one may introduce
variants of trimaximal mixing, such as  
\be \label{eq:TMv2}
\mbox{TM}_1:~~
\left(
\ba 
|U_{e1}|^2 \\
|U_{\mu 1}|^2 \\
|U_{\tau 1}|^2
\ea
\right) = 
\left(
\ba 2/3 \\ 1/6 \\ 1/6
\ea
\right) \,.
\ee
Here we have fixed the first column of $U_{\rm PMNS}$ 
to have the same form as in the case of tri-bimaximal 
mixing\footnote{This 
possibility has been mentioned first in Ref.~\ \cite{Lam}, 
but its phenomenology has not been studied yet.}. 
A consequence of Eq.~(\ref{eq:TMv2}) is that $\sin^2 \theta_{12} 
\le \frac 13$. Indeed, from $|U_{e1}|^2 = \frac 23$ one finds 
\be
\sin^2 \theta_{12} = \frac 13 \, \frac{1 - 3 \, |U_{e3}|^2}
{1 - |U_{e3}|^2} \simeq 
\frac 13 \, \left( 1 - 2 \, |U_{e3}|^2 \right) \,.
\ee
Using the range $|U_{e3}|^2 = 0.016 \pm 0.010$ gives 
$\sin^2 \theta_{12} = 0.322 \pm 0.007$. Fig.~\ref{fig:TMt12a} shows 
$\sin^2 \theta_{12}$ as a function of $|U_{e3}|$. In contrast to the
original trimaximal mixing scheme, TM$_2$, the best-fit value 
of $\sin^2 \theta_{12}$ can be obtained (for $|U_{e3}| \simeq
0.179$). 
Note that $\sin^2 \theta_{12}$ decreases with 
$|U_{e3}|$. As long as $|U_{e3}| \gs 0.06$, $\sin^2 \theta_{12}$ is
within its allowed $1\sigma$ range. 

The second independent condition in Eq.~(\ref{eq:TMv2}) involving 
$|U_{\mu 1}|^2 = 1/6$ gives 
\be
\cos \delta \, \tan 2 \theta_{23} = 
 - \frac{1 - 5 \, |U_{e3}|^2}{2\sqrt{2} \, |U_{e3}| \, 
\sqrt{1 - 3 \, |U_{e3}|^2}} \\
\simeq 
\frac{-1}{2\sqrt{2} \, |U_{e3}|} \left(1 - \frac{7}{2} \, 
|U_{e3}|^2 \right)\, . 
\ee
Here the results are qualitatively similar to the ones for 
trimaximal mixing treated in Sect.~\ref{sec:TM}.
In Fig.~\ref{fig:TMt12a} we also show $\sin^2 \theta_{23}$ 
as a function of $\delta$. As for TM$_2$ mixing, 
CP conserving values of $\delta$ maximize the deviations, 
but here the $3\sigma$ range can easily be overshot. 
Finally, in Fig.~\ref{fig:TMt12a}, $\sin^2 \theta_{23}$ is plotted 
against $\sin^2 \theta_{12}$. Again, the possible departure from 
$\sin^2 \theta_{23} = \frac 12$ can be larger than the one from 
$\sin^2 \theta_{12} = \frac 13$. 

A mixing matrix with the ${\rm TM}_1$ property can be 
obtained by multiplying $U_{\rm TBM}$ with a 23-rotation of 
angle $\theta$ from the right, 
$U_{\rm TM_1} = U_{\rm TBM} \, R_{23}(\theta; \psi)$. 
The observables are in this case 
\bea \D 
|U_{e3}|^2 = \frac 13 \, \sin^2 \theta
~,~~\sin^2 \theta_{23} = \frac 12 - \frac{\sqrt{\frac 32} \, \sin
2\theta \, \cos \psi }{3 - \sin^2 \theta}~,~~\\ \D 
\sin^2 \theta_{12} = 1 - \frac{2}{3 - \sin^2 \theta}~,~~
J_{\rm CP} = \frac{1}{6\sqrt{6}} \, \sin 2 \theta 
\, \sin \psi\,,
\eea
where the CP phase $\delta$ is again related to $\psi$. 
For the mass matrix we find 
\[ 
\label{eq:TM1matrix}
  (m_\nu)_{\rm TM_1} = 
U^\ast_{\rm TM_1} \, P^\ast \, m^{\rm diag}_\nu \, 
      P^\dagger \, U^\dagger_{\rm TM_1}
    = \left( \begin{array}{ccc} A & B + C & B - C\\ 
    \cdot & \frac{1}{2} \, (A + B + D + 4 \,C) 
& \frac{1}{2} \, (A + B - D)\\
     \cdot & \cdot & \frac{1}{2} \, (A + B + D - 4 \, C)\\
      \end{array}\right),
\]
where we identify 
\[ 
\label{eq:TM1elem}
\begin{array}{rl}
  A = \frac 13 \left( 2 \, m_1 + m_2 \, c^2_\theta \, e^{-2i\alpha} 
        + m_3 \, s_\theta^2 \, e^{2i(\psi - \beta)}\right)\, ,\qquad&
  C = \frac{1}{\sqrt{6}} \left(m_2 e^{-i(\psi+2\alpha)}- m_3 \, 
        e^{i(\psi-2\beta)} \right) s_\theta \, c_\theta\,,\\
  B = \frac 13  \left( -m_1 + m_2 \, c^2_\theta \, e^{-2i\alpha} +
        m_3 \, s^2_\theta\, e^{2i(\psi - \beta)} \right)\,,\qquad&
  D = m_2 \, e^{-2i(\psi+\alpha)} \, s^2_\theta 
+ m_3 \, c_\theta^2 \, 
        e^{-2i\beta}\, .\\
\end{array}
\]
Again we see that the original $\mu$--$\tau$ symmetry is 
broken by the extra terms involving $C$. We can decompose 
the mass matrix for TM$_1$ in terms of the individual neutrino 
masses and find that $m_1$ is multiplied with the same matrix as 
in Eq.~(\ref{eq:mnuind}).  

If we would insist that the third column of 
$U_{\rm TBM}$ remain invariant instead, i.e., 
$|U_{e3}|^2 = 0$, $|U_{\mu3}|^2 = |U_{\tau3}|^2 = \frac 12$, 
then $\theta_{13} = \delta = 0$, 
$\theta_{23} = \pi/4$, while $\theta_{12}$ is 
a free parameter. This case (${\rm TM_3}$ in our notation) 
is nothing other than the well-known $\mu$--$\tau$ symmetry.\\ 

It is also of interest to consider the case where one of 
the rows of the tri-bimaximal mixing matrix remains invariant. 
Again such a result can be obtained by multiplying 
with a suitable two dimensional rotation 
matrix, but now from the left. 
In fact, this class of deviations from 
tri-bimaximal mixing corresponds to the charged lepton 
flavor matrix differing from its diagonal mass matrix, thus 
introducing a $U^\dagger_\ell$ factor in the PMNS mixing 
matrix, i.e., $U_{\rm PMNS} = U^\dagger_\ell \, U_{\rm TBM}$ 
\cite{CL}.

Let us start with the case of the first row in 
$U_{\rm TBM}$ remaining invariant. We denote this with 
a superscript as 
\be
\mbox{TM$^1$}:\qquad 
\left( 
|U_{e 1}|^2 \,,~|U_{e 2}|^2\,,~|U_{e 3}|^2 \right) = 
\left( 
\frac 23\,,~\frac 13 \,,~0
\right)\,,
\ee 
As a result, $\theta_{23}$ is a free parameter, while 
$\sin^2 \theta_{12} = \frac 13$, as well as $\theta_{13} = \delta =
0$. With a rotation of $U_{\rm TBM}$ by the matrix 
$R_{23}(\theta; \psi)$ from the left, the light 
Majorana neutrino mass matrix becomes 
\be
\label{eq:TM^1}
  (m_\nu)_{\rm TM^1} = R_{23}^\ast \, 
(m_\nu)_{\rm TBM} \, R^\dagger_{23} \, .
\ee

\begin{figure}[ht]
\begin{center}
\epsfig{file=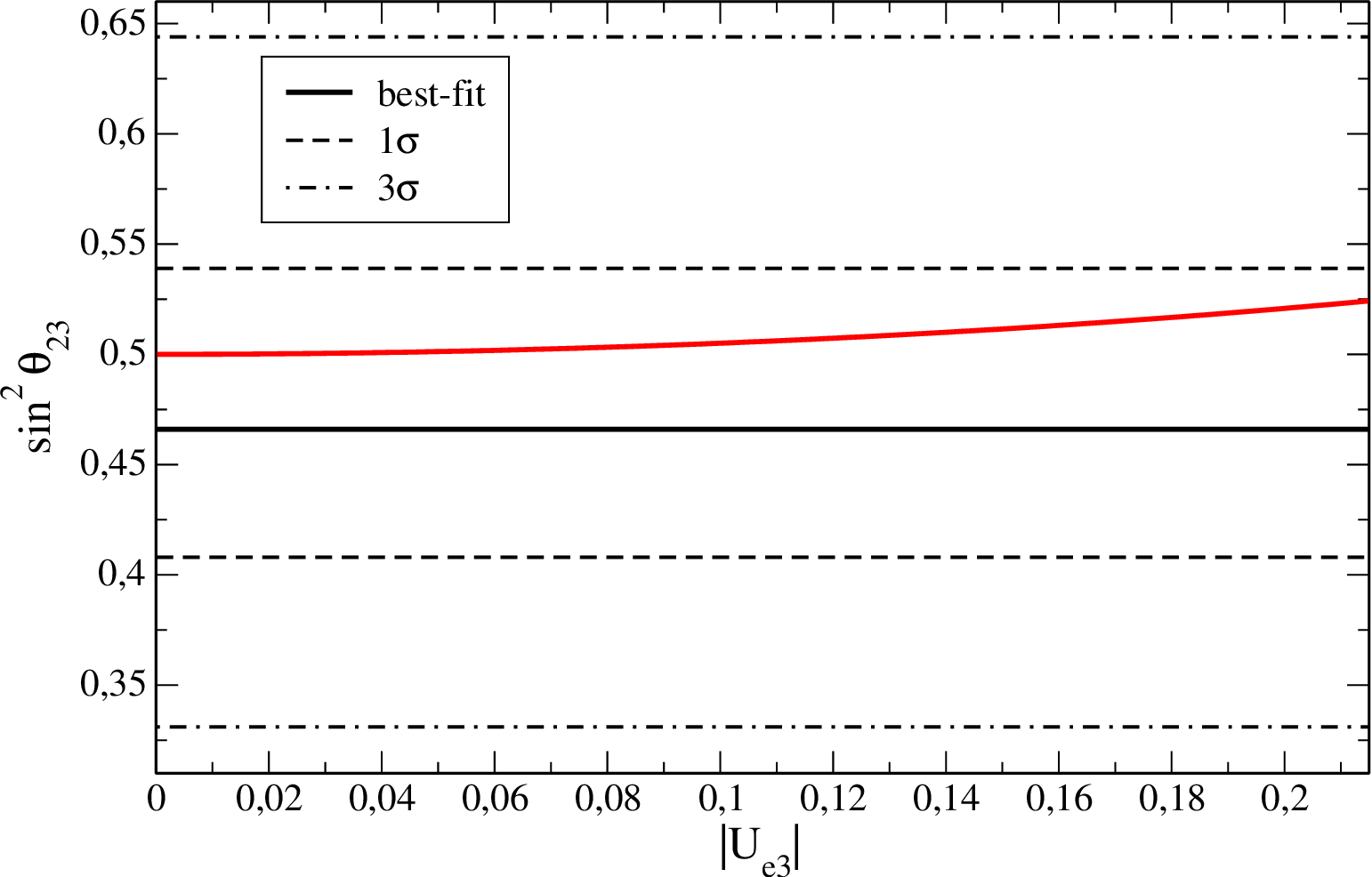,width=10cm,height=7cm}
\epsfig{file=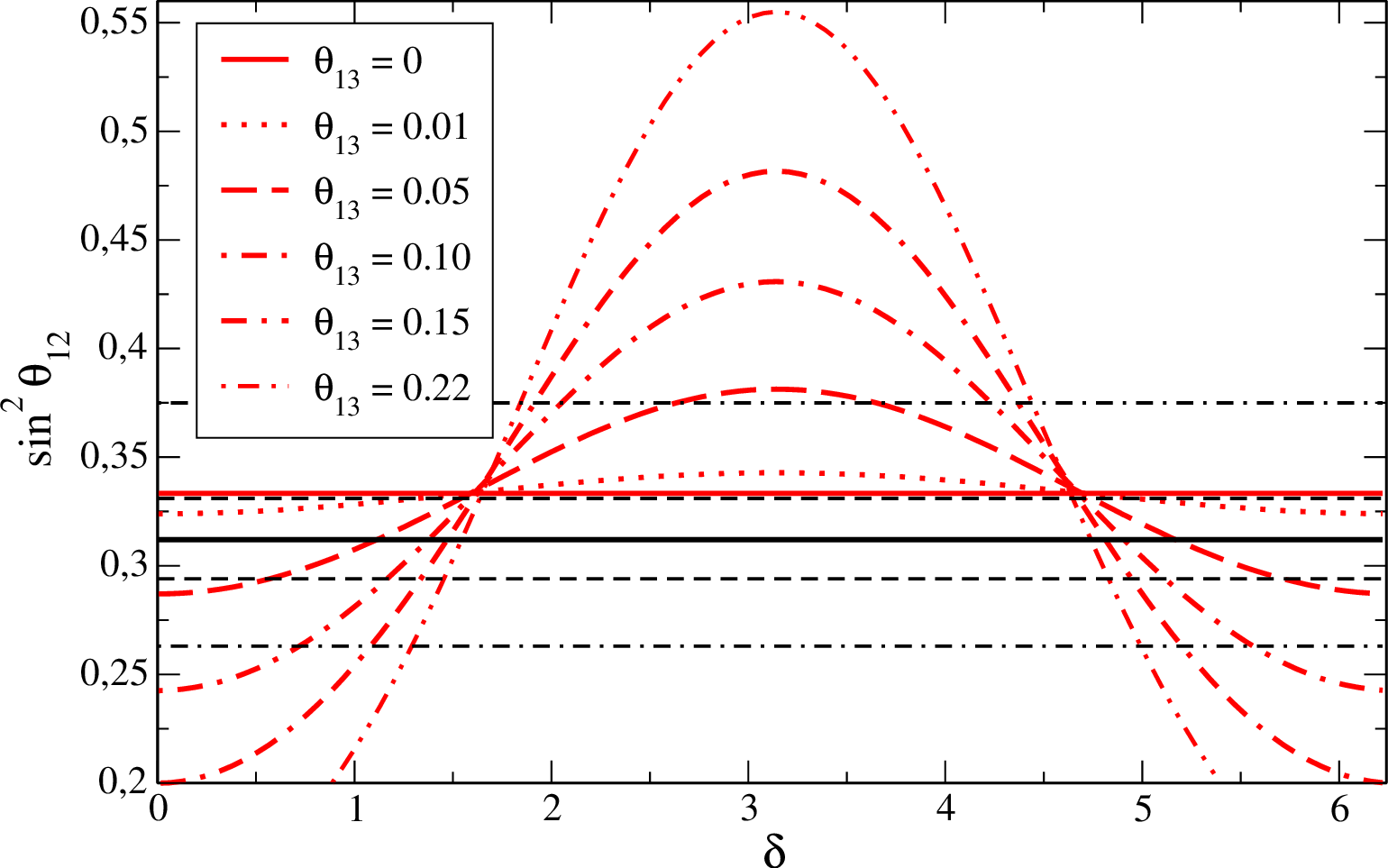,width=10cm,height=7cm}
\epsfig{file=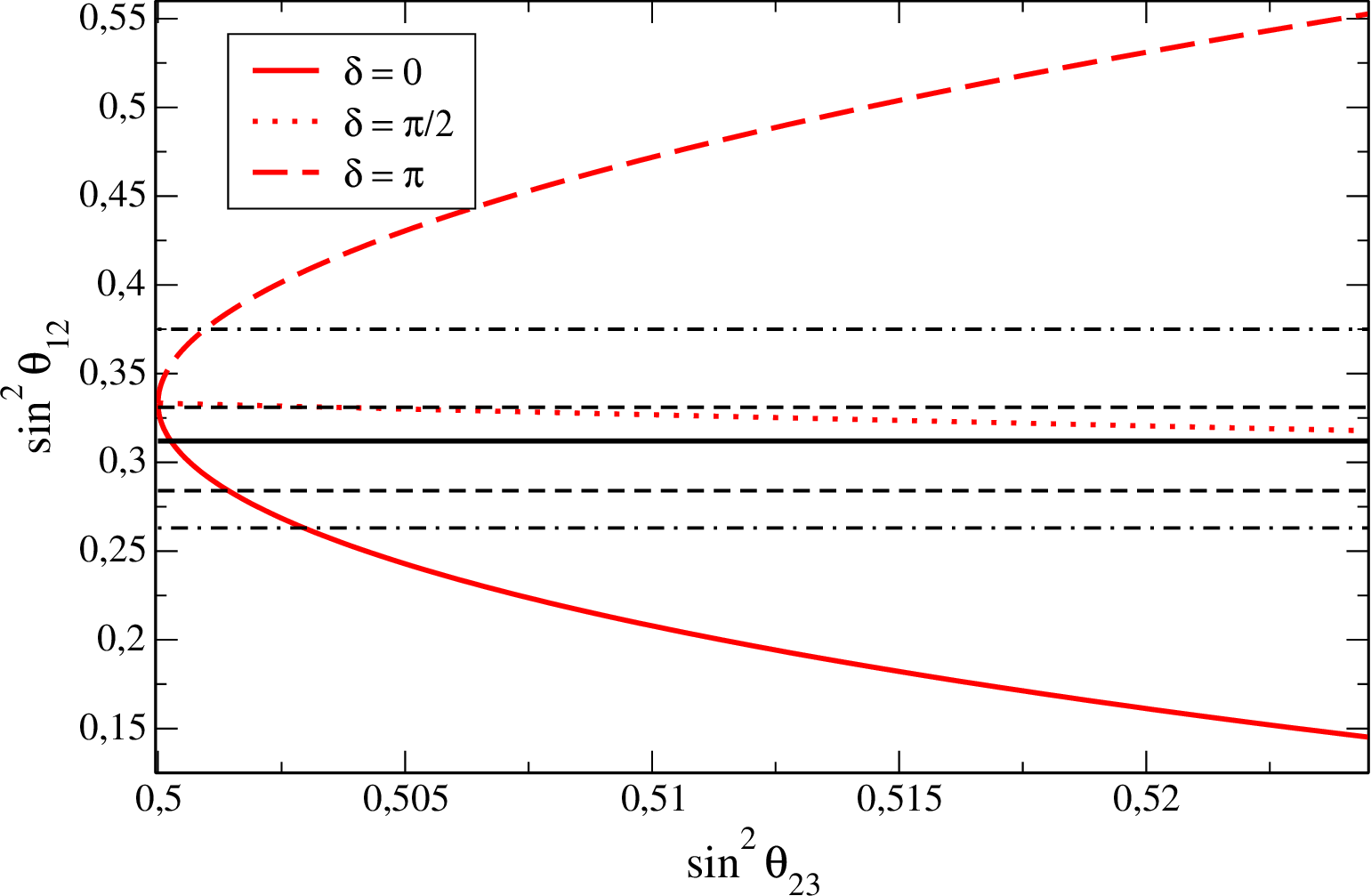,width=10cm,height=7cm}
\caption{\label{fig:TMt12b}
Phenomenology of exact TM$^2$ mixing. 
Shown are the atmospheric neutrino parameter $\sin^2
\theta_{23}$ against $|U_{e3}|$, the solar neutrino 
parameter $\sin^2 \theta_{12}$ against $\delta$ for 
different values of $|U_{e3}|$ and $\sin^2
\theta_{12}$ against $\sin^2 \theta_{23}$. 
Also given are the current best-fit value and the $1\sigma$ as well as
$3\sigma$ ranges from global fit \cite{bari}. } 
\end{center}
\end{figure}

If we consider the second or third row we can correlate all 
four mixing parameters. Starting with the second row, i.e., 
\be
\mbox{TM}^2: \qquad 
\left( 
|U_{\mu 1}|^2 \,,~|U_{\mu 2}|^2\,,~|U_{\mu 3}|^2 
\right) = 
\left( 
\frac 16\,,~\frac 13 \,,~\frac 12
\right)\,,
\ee 
one immediately finds from $|U_{\mu 3}|^2 = \frac 12$: 
\be \label{eq:s23v3}
\sin^2 \theta_{23} = \frac{1}{2 \, (1 - |U_{e3}|^2)} 
\simeq \frac 12 \left(1 + |U_{e3}|^2 
\right) \ge \frac 12\,,
\ee
i.e., atmospheric neutrino mixing on the ``dark side,'' 
with a maximal value of $\sin^2 \theta_{23} \simeq 0.524$ for 
$|U_{e3}|^2 = 0.046$. 
Inserting Eq.~(\ref{eq:s23v3}) in $|U_{\mu 2}|^2 = \frac 13$ 
gives a complicated and lengthy expression including $\cos \delta$, 
$\sin^ 2\theta_{12}$ and $|U_{e3}|$, which can be approximated as 
\be
\sin^2 \theta_{12} \simeq \frac 13 
- \frac{2\sqrt{2}}{3} \,|U_{e3}|\, 
\cos \delta \,  + \frac 13 \, |U_{e3}|^2 \,\cos 2 \delta \,.
\ee
Fig.~\ref{fig:TMt12b} shows the result for the TM$^2$ scenario. 
The parameter dependence is similar to the one for 
the TM$_1$ and TM$_2$ scenarios, the main difference being the 
exchanged roles of $\theta_{12}$ and $\theta_{23}$.  

\begin{figure}[ht]
\begin{center}
\epsfig{file=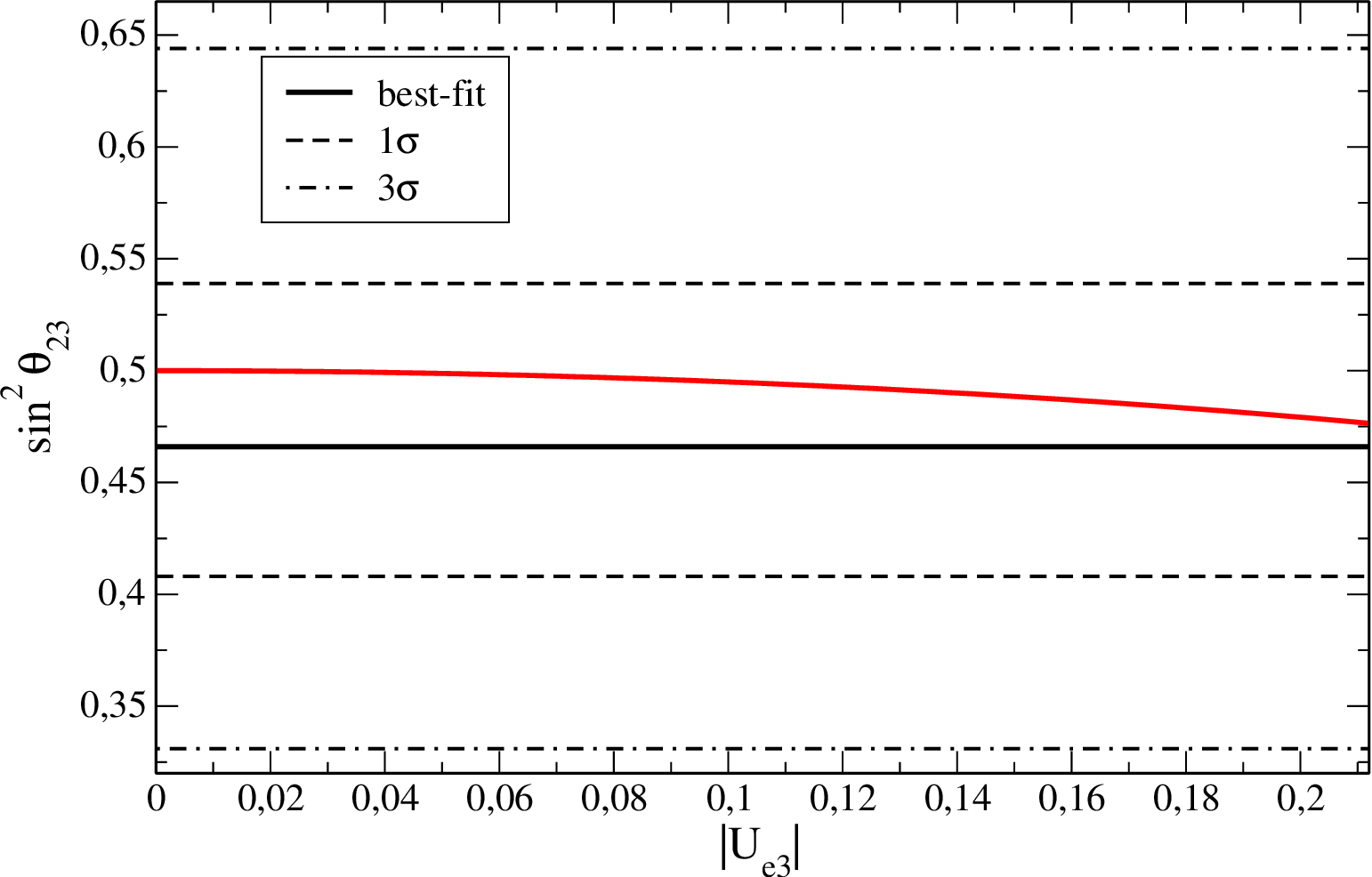,width=10cm,height=7cm}
\epsfig{file=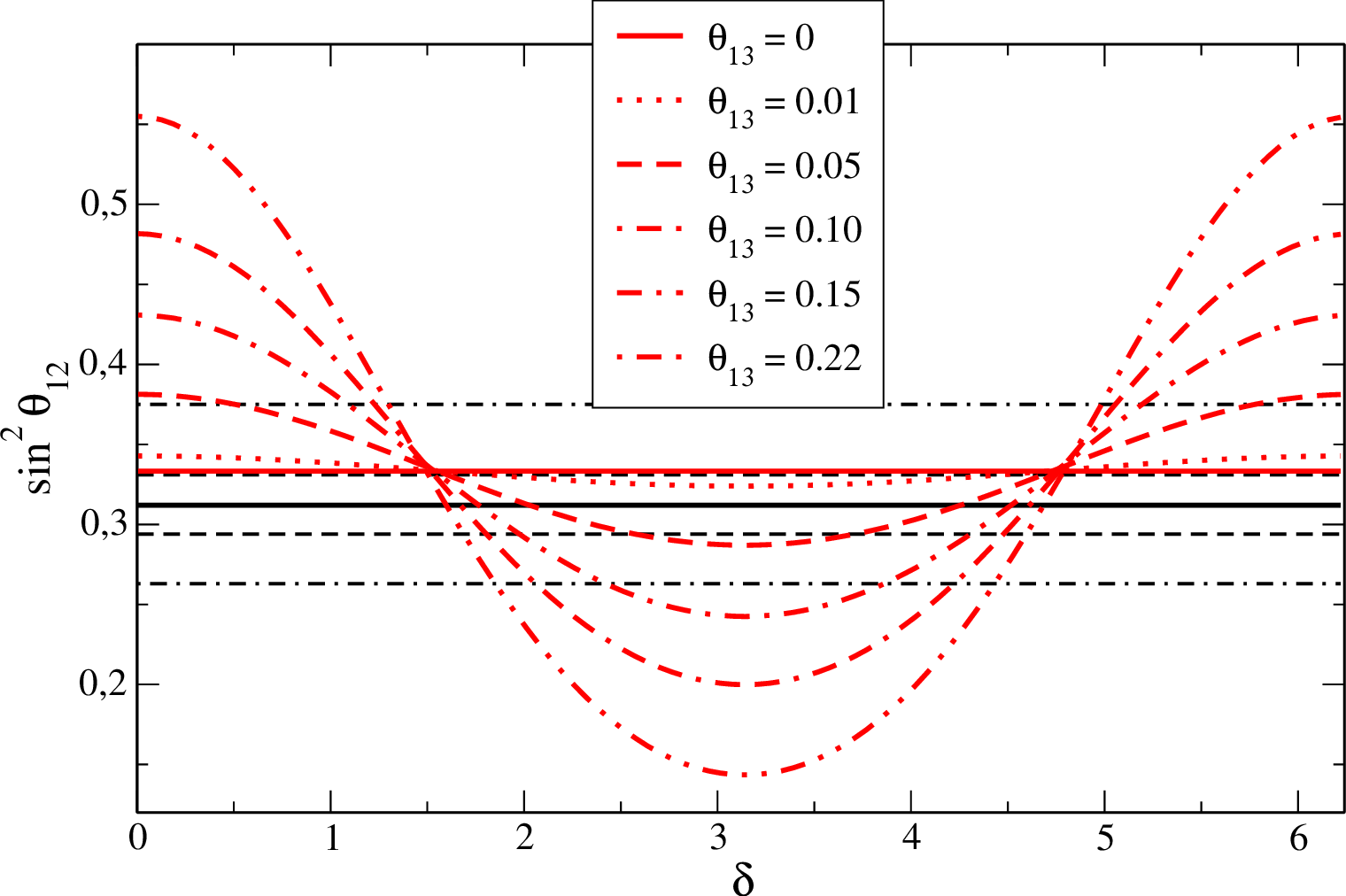,width=10cm,height=7cm}
\epsfig{file=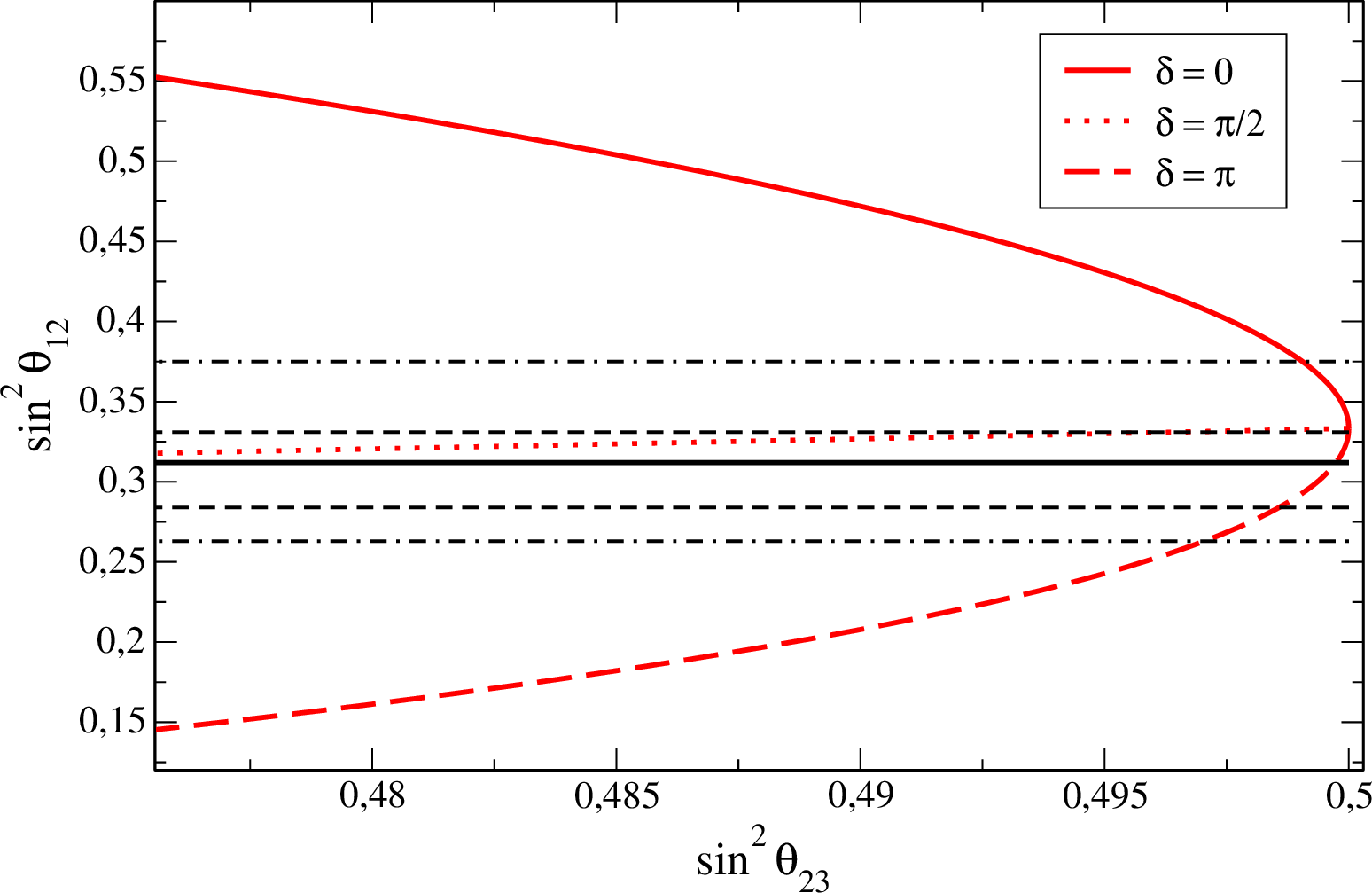,width=10cm,height=7cm}
\caption{\label{fig:TMt12c}
Phenomenology of exact TM$^3$ mixing. 
Shown are the atmospheric neutrino parameter $\sin^2
\theta_{23}$ against $|U_{e3}|$, the solar neutrino 
parameter $\sin^2 \theta_{12}$ against $\delta$ for 
different values of $|U_{e3}|$ and $\sin^2
\theta_{12}$ against $\sin^2 \theta_{23}$. 
Also given are the current best-fit value and the $1\sigma$ as well as
$3\sigma$ ranges from a global fit \cite{bari}. } 
\end{center}
\end{figure}

On the other hand with the third row remaining invariant under the 
transformation 
\be
\mbox{TM}^3:~
\left( 
|U_{\tau 1}|^2 \,,~|U_{\tau 2}|^2\,,~|U_{\tau 3}|^2 
\right) = 
\left( 
\frac 16\,,~\frac 13 \,,~\frac 12
\right)
\ee 
the same approximate formula with a relative sign for the term 
of order $|U_{e3}|$ is found. Now, however, one finds 
atmospheric neutrino mixing on the ``bright side'': 
\be
\sin^2 \theta_{23} = \frac{1 - 2\, |U_{e3}|^2}{2 \, (1 - 
|U_{e3}|^2)} \simeq \frac 12 \left(1 - |U_{e3}|^2 
\right) \le \frac 12\,.
\ee 
The maximal deviation occurs for the
largest possible $|U_{e3}|^2 = 0.046$, in which case 
$\sin^2 \theta_{23} \simeq 0.476$. 
In Fig.~\ref{fig:TMt12c} the resulting correlations 
for the TM$^3$ scenario are given. 

As in the case of $\rm TM^1$, the $U_{\rm TM^2}$ mixing 
matrix can be obtained by a rotation of $U_{\rm TBM}$ from 
the left by $R_{13}$.  For $\rm TM^3$, the 
rotation matrix is $R_{12}$. For all three cases with the 
first, second, or third row of $U_{\rm TBM}$ remaining invariant, 
solar neutrino mixing can occur in the complete $3\sigma$ range. 
The deviations from tri-bimaximal mixing can be considerably 
larger for $\theta_{13}$ in the normal hierarchical
case than were observed in Sect.~\ref{sec:AR} 
and Ref.~\cite{AR}.\\

Finally, it is amusing to compare the trimaximal mixing schemes 
with the recently proposed tetramaximal one \cite{tetra}. 
Its name stems from the fact that it can be obtained by four 
consecutive rotations each having a maximal angle of $\pi/4$, 
and with appropriately chosen phases: 
$U_{\rm tetra} = R_{23} (\pi/4; \pi/2) \, R_{13} (\pi/4; 0) \, 
R_{12} (\pi/4; 0) \, R_{13} (\pi/4; \pi)$. The predictions are 
$\delta= \pi/2$, $\sin^2 \theta_{23} = \frac 12$, 
$|U_{e3}|^2 = \frac 14 \, (\frac 32 - \sqrt{2}) \simeq 0.021$ 
and $\sin^2 \theta_{12} = 1/(\frac 52 + \sqrt{2}) 
\simeq 0.255$. None of the trimaximal variants discussed in this 
paper could be confused with tetramaximal mixing.

\section{\label{sec:concl}Summary and Conclusions}

Although the present PMNS lepton mixing matrix deduced 
from experiment is consistent with tri-bimaximal mixing, 
it is of interest to study possible deviations which 
may arise in the future. In a previous paper the authors 
considered linear complex perturbations of the neutrino 
matrix away from the most general $\mu$--$\tau$ symmetric 
texture which yields tri-bimaximal mixing. 
Here we have considered variations of trimaximal mixing which can
arise, e.g., with a simple complex rotation of 
the $U_{\rm TBM}$ matrix from the right
or the left. 
The original trimaximal mixing matrix $U_{\rm TM_2}$ preserves
the second column of $U_{\rm TBM}$ for which each 
element has absolute value of $1/\sqrt{3}$. 
We have generalized this mixing scenario here and 
refer to the other variations 
as TM$_k$ or TM$^k$ according to which $k = 1,2,3$ column or 
row remains invariant, respectively. 
Independent of our comparisons with tri-bimaximal mixing, 
the ``trimaximal'' mixing scenarios considered here are alternative, 
novel and testable mixing schemes. 
 
\begin{table}[ht]
\begin{center}
\begin{tabular}{lccc}
\hline\hline\\
    &  $|U_{e3}|^2$    & $\sin^2 \theta_{12}$ &  $\sin^2 \theta_{23}$\\[0.1in]
\hline\\
TBM    &       0	      &    $\frac13$       &    
$\frac 12$ \\[0.1in]
Broken TBM &  \\
\qquad  NH       &  \lsim 0.001  &   $0.26 - 0.38$   &  
$0.37 - 0.63$ \\
\qquad  IH       &  \lsim 0.014     &    $0.26 - 0.38$   &  
$0.35 - 0.64$ \\[0.1in]
${\rm TM}_1$ &   $\le 0.046$   &  $ \frac{1}{3} \, 
(1 - 2 \, |U_{e3}|^2)$ & 
      $\frac{1}{2} - \sqrt{2} \, |U_{e3}| \, 
(1 - \frac{1}{2} \, |U_{e3}|^2)
          \cos \delta$ \\
	&    &   $0.30 - 0.33$  &   $0.34 - 0.64$ \\[0.1in]
${\rm TM}_2$ &  $\le 0.046$   & $\frac{1}{3} \, (1 + |U_{e3}|^2)$ & 
    $\frac{1}{2} + \frac{1}{\sqrt{2}}  \, |U_{e3}| \, (1 + 
	\frac{1}{4} \, |U_{e3}|^2) \, \cos \delta$ \\
	&     &  $0.33 - 0.35$  &   $0.34 - 0.64$ \\[0.1in]
${\rm TM}_3$ &  0  &  $0.26 - 0.38$  &   $\frac 12$ \\[0.1in] 
${\rm TM}^1$ &  0  &   $\frac 13$   &   $0.34 - 0.64$ \\[0.1in]
${\rm TM}^2$ &  $\le 0.046$  & $\frac{1}{3} - 
\frac{2\sqrt{2}}{3}  \, |U_{e3}| \, 
	\cos \delta + \frac{1}{3} \, |U_{e3}|^2  \, \cos 2\delta$
	& $\frac{1}{2} \, (1 + |U_{e3}|^2)$\\
	&    &   $0.26 - 0.38$  & $0.50 - 0.52$ \\[0.1in]
${\rm TM}^3$ &  $\le 0.046$  & $\frac{1}{3} 
+ \frac{2\sqrt{2}}{3} \, |U_{e3}|  \, 
	\cos \delta + \frac{1}{3} \, |U_{e3}|^2  \, \cos 2\delta$ 
	& $\frac{1}{2} \, (1 - |U_{e3}^2)$\\ 
	&    &  $0.26 - 0.38$   &  $0.49 - 0.50$ \\
\hline\hline
\end{tabular}
\caption{\label{Table}Summary of the mixing angles 
obtained in the  exact and broken tri-bimaximal, and 
generalized trimaximal mixing schemes. We have for the 
sake of illustration expanded the exact correlations 
(see Figs.~\ref{fig:TMt12} -- \ref{fig:TMt12c}). 
For the ranges given, the present 3$\sigma$ data bounds on 
the three mixing angles have been imposed.}
\end{center}
\end{table}

\begin{figure}[ht]
\begin{center}
\epsfig{file=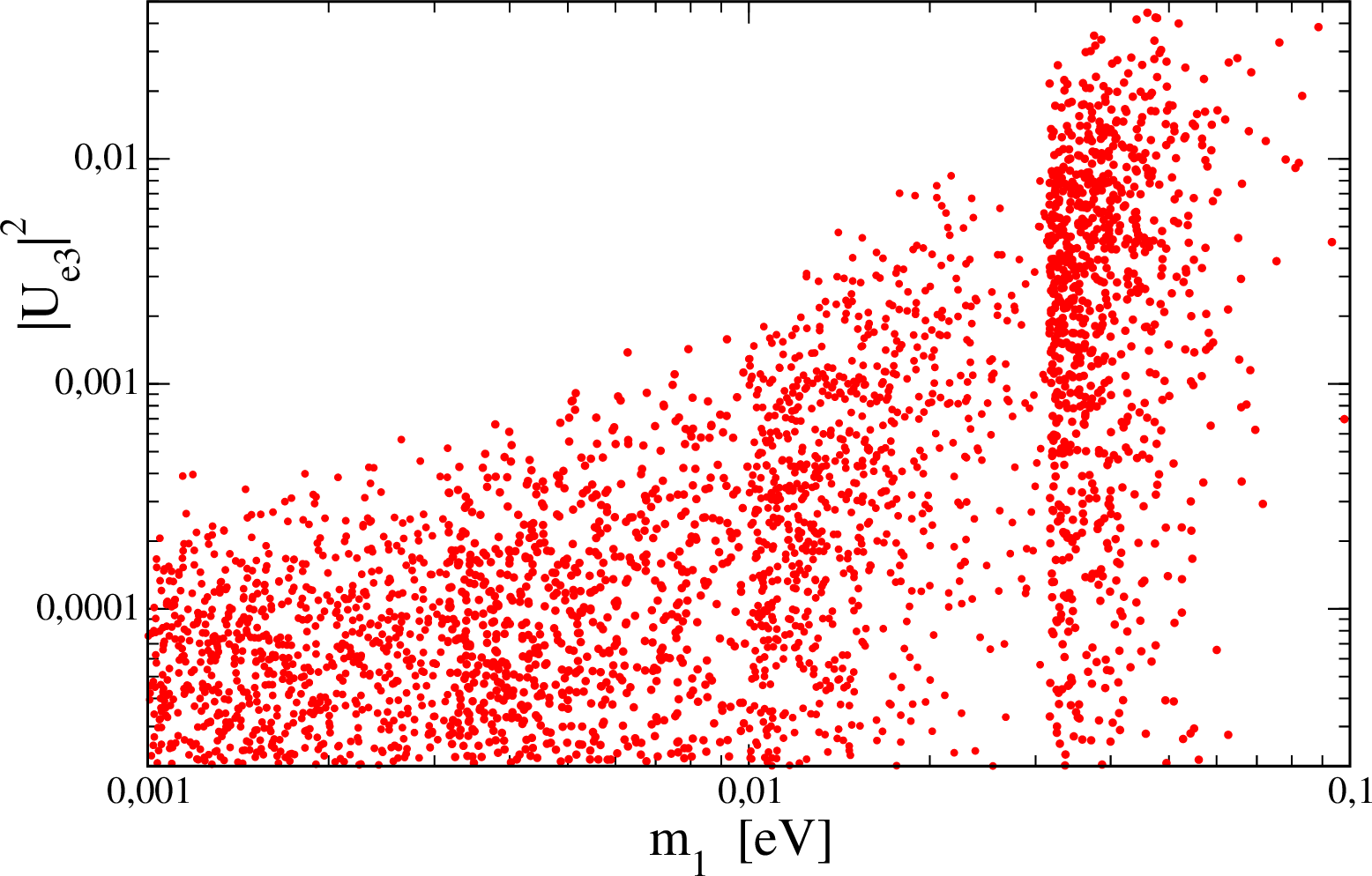,width=10cm,height=7cm}
\epsfig{file=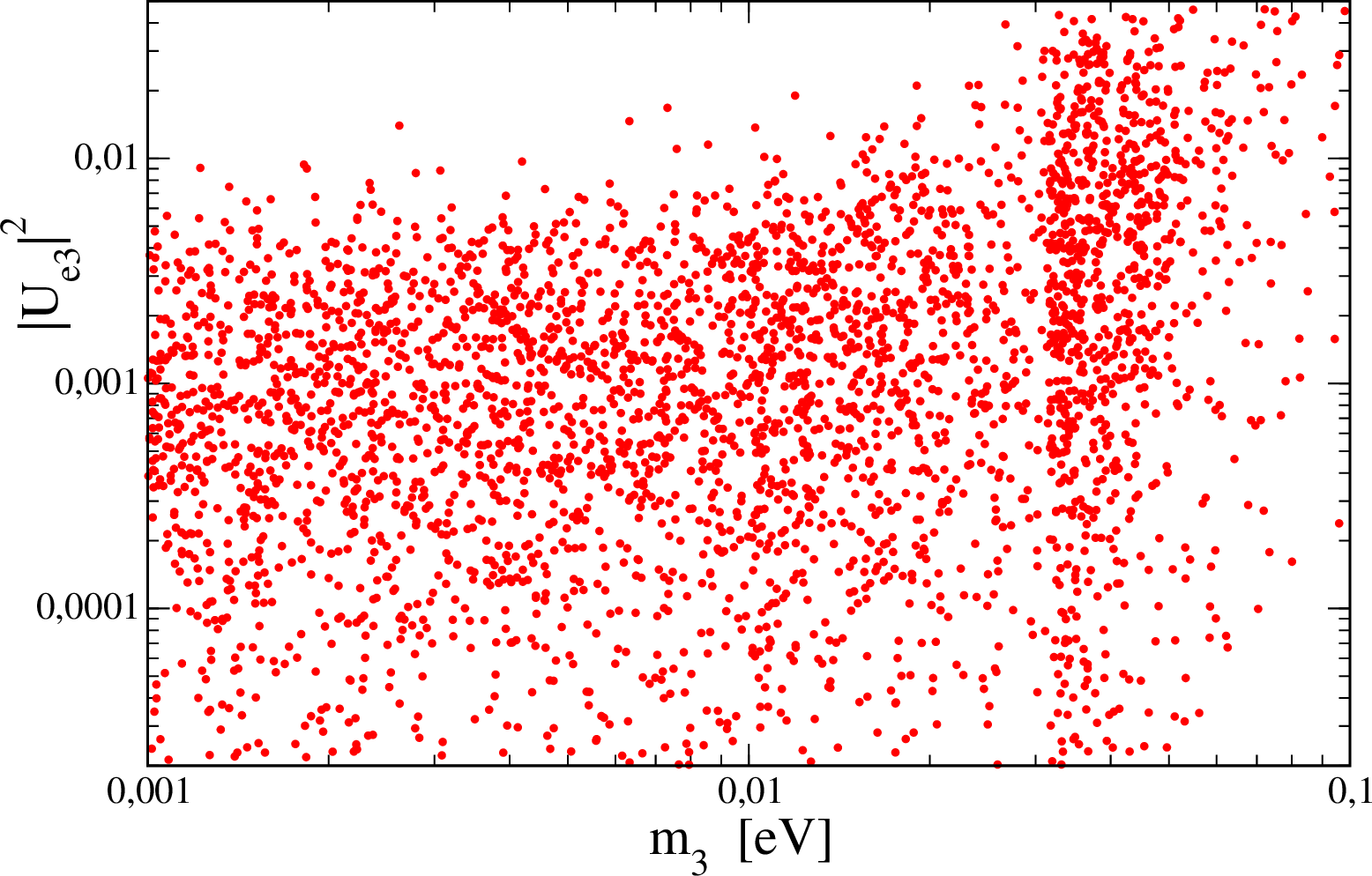,width=10cm,height=7cm}
\caption{\label{fig:m1}Scatter plot of the predictions for
$|U_{e3}|^2$ versus the smallest neutrino mass for softly broken 
tri-bimaximal mixing. Shown are the normal mass ordering (upper plot)
and the inverted one (lower plot).} 
\end{center}
\end{figure}

We summarize our findings in Table \ref{Table} 
for the squares of the sines of the three mixing angles. 
For broken TBM where up to 20\% deviations are
allowed in every neutrino mass matrix element, 
we had found that the allowed variations in sine 
squared of the mixing angles are uncorrelated to a large extent. 
The perturbed solar and atmospheric neutrino mixing angles 
cover the entire mixing ranges presently allowed, 
while $|U_{e3}|^2 = \sin^2 \theta_{13}$ can 
range from zero up to 0.001 (0.014) for the case 
of normal (inverted) hierarchy.  These upper bounds 
depend sensitively on the lightest neutrino 
mass as shown in Fig.~\ref{fig:m1}, 
especially for a normal ordering.  The largest 
upper bounds presently allowed are reached in the case of 
three-fold neutrino mass degeneracy.

For the six generalized trimaximal mixing cases considered, 
on the other hand, the mixing angles and Dirac CP phase are 
characteristically correlated for four of the cases. 
The exceptional cases arise for ${\rm TM_3}$ and ${\rm TM^1}$.  
For ${\rm TM_3}$ both $|U_{e3}|^2$ and 
$\sin^2 \theta_{23}$ remain fixed at their TBM values, 
while $\theta_{12}$ is a free variable and limited only 
by the present experimental bounds. This case corresponds to the 
well-known $\mu$--$\tau$ symmetry. 
For ${\rm TM^1}$, $|U_{e3}|^2$ and $\sin^2 \theta_{12}$ 
remain fixed, while $\sin^2 \theta_{23}$ remains 
bounded only by experiment. 
For ${\rm TM_1}$ and ${\rm TM_2}$ the solar mixing 
is tightly limited, with the former ranging just below 
the TBM value and the latter just above the TBM value 
of $\frac{1}{3}$.  The atmospheric neutrino mixing in these 
two cases can cover the full presently allowed 
region with the former (latter) peaking at 
$\delta = 0 \, (\pi)$ and bottoming at $\delta = \pi \, (0)$. 
For ${\rm TM^2}$ and ${\rm TM^3}$ the opposite situation holds, 
where the atmospheric neutrino mixing range is tightly
limited close to the TBM value in the bright side or dark side, 
respectively, while the full range for solar 
neutrino mixing can be realized.  

When more refined ranges for the mixing angles are known, 
one will be able to rule out or confirm the new mixing scenarios 
discussed here and narrow down the acceptable 
deviations from tri-bimaximal mixing we have 
discussed in this paper. 

\vspace{0.3cm}
\begin{center}
{\bf Acknowledgments}
\end{center}

W.R.~wishes to thank Walter Grimus and 
the Universit\"at Wien, where parts of this 
work were carried out, for kind hospitality and discussions. 
This work was supported by the ERC under the Starting Grant 
MANITOP and by the Deutsche Forschungsgemeinschaft 
in the Transregio 27 ``Neutrinos and beyond -- weakly interacting 
particles in physics, astrophysics and cosmology'' (W.R.). 
C.H.A.~thanks the members of the Fermilab Theory Group for their 
kind hospitality.

\newpage

\pagestyle{empty}

\end{document}